# The Prospects for e-Examinations in Nigeria and Australia

Olawale S. Adebayo and Shafi'i M. Abdulhamid
Department of Cyber Security Science
School of Information and Communication Technology
Federal University of Technology
Minna, Nigeria
(waleadebayo, shafii.abdulhamid)@futminna.edu.ng

Andrew Fluck
Faculty of Education
University of Tasmania
Launceston, Australia
Andrew.Fluck@utas.edu.au

*Abstract* :

**This paper compares the e-Examination system in Nigeria with that of Australia. We consider the experiences of working with commercial firms such as Electronic Testing Company (eTC) and using open-source software. It is important to foster good relationships with accreditation authorities (such as University Authorities, West African Examination Council (WAEC), Joint Admissions and Matriculation Board (JAMB) etc. and the Tasmanian Qualifications Authority) to assist in the transition from paper-based assessment to post-paper assessment. The paper also considers the relative convenience for students, administrators and lecturer/assessors; and to gauges the reliability and security of the two systems in use. It examines the challenges in conducting e-Examinations in both countries by juxtaposing the systems in the two countries and suggests ways of developing more acceptable e-Examination systems.**

## I. Introduction

Good assessment is integral to good teaching. Good teaching provides direction for future generations, but is also influenced by political pressures and is responsive to social needs. Because technology is changing so quickly (the internet started in 1985 and is now available almost everywhere; certainly in hotels in the central pacific islands of Kiribati) social linkages with curriculum have come under strain as common usage of computers diverges from teaching norms in schools.

This is the context into which the authors have pioneered the use of eExaminations in their different countries. By eExaminations we mean the use of computers by candidates in a high stakes supervised assessment generally occurring simultaneously over a fixed length of time. There are many other uses of computers in learning, and many other kinds of possible assessment. Learning content management systems are common for delivering learning materials over the world-wide-web and facilitating dispersed students. Assessing such learning using pen-on-paper examinations can appear incongruous to the students – but it is often encountered. Coursework assessment, where students take a problem home for completion over a number of days is also common. Good assessment will often combine this kind of approach with more formal examinations. In this paper we make no defence of the choice of the examination strategy of assessment; that is left to accrediting authorities. However, this paper will focus upon this one style of assessment, since it is the focus of our endeavors and often crucial for university entry and progression in university undergraduate courses.

## II. LITERATURE

### A. eExaminsations in the secondary (high school) sector

Some trials of the necessary technologies have been conducted in secondary schools. In 2012 the University of Cambridge International Examinations board provided Impington Village College candidates with Kindles and iPads loaded with the test papers, although students still had to handwrite the answers in a mock IGCSE biology exam. Whilst the majority favoured the use of technology, one third of the students preferred pen-on-paper [1], [2]. In the USA, On December 20, 2010, the US Department of Education issued the Assessment Technology Standards Request for Information [3]. This request and responses outline the range of technical matters such as compatibility standards required for widespread adoption of any new technology used in high stakes assessment. Of particular interest are the significant considerations around internet or network access of any kind, and the reliability of the equipment.

### B. eExaminations for university courses

Some universities have explored the use of technology for high stakes examinations. The arguments for various technical methods have been discussed elsewhere [4] and security methods compared [5]. Some 86% of Law schools in the USA and thirty three others elsewhere have investigated or adopted a computer-based approach to essay-style examinations.

Where candidates have a free choice of medium for answering an examination, initial reports indicate there are no systematic differences in achievement levels [6]. More recent evidence suggests that computer-using candidates have a slight advantage, with greater word-counts and more complex language [7].

Our universities have considered the use of computers at the highest level. In February, 2010 the Federal University of Technology, Minna in Nigeria approved the use of computers in examinations, and on 4th March 2011 the University of Tasmania (UTAS) Academic Senate also approved the use of eExams (in all disciplines). To make these decisions, consideration was given to technical reliability, equity and implementation processes. [8] reviewed the eExams systems in FUT Minna and some other Universities in Nigeria and designed a new system which uses data encryption in order to protect the questions sent to the e-Examination center through the internet or intranet and a biometric fingerprint authentication to screen the stakeholders.

*Andrew Fluck visited FUT Minna whilst on paid long service leave from the University of Tasmania. We wish to acknowledge the efforts and understanding of the eTC manager, FUT Minna for conducting us round their facilities for verification and analysis.*

III. METHODOLOGY

The current study was undertaken subsequent to a visit to FUT Minna, and uses a comparative case study design. Sample selection was by opportunity, since one author had become aware of the work at FUT Minna, and arranged to visit the campus, on the basis of searches using the 'eExamination' term. It is not intended these case studies should have similarities; in fact the intentions and technologies at the two sites are very diverse, so give a greater perspective on the subsequent issues than a closer match might have revealed.

Yin defines case study research method as an empirical inquiry that investigates a contemporary phenomenon within its real-life context and in which multiple sources of evidence are used [9]. Our data were collected holistically from the individual sites, and then organized into high level categories for presentation and subsequent analysis.

Bias in the face of limited samples and selection bias are the obvious threats to validity in this approach [10]. In this paper the two cases represent markedly different populations, education systems and political contexts, therefore selection bias is considered minimal. Our results will have limited validity, but we expect to show *whether* and *how* eExaminations can play a part in selecting students for university courses, not to what extent this is true.

The case studies are provided in the following section, with that from FUT Minna first, and from the University of Tasmania afterwards. Each case study is arranged in four parts, with a general introduction which defines the technological approach and educational context; use in university entrance selection processes; use in university courses; and expected future directions.

IV. CASE STUDY ONE: FUT MINNA

A. *Background*

The Electronic Testing Company (eTC) is based in Lagos, Nigeria and entered into a public-private partnership with the Federal University of Technology Minna in 2010 with an agreement to build a standard assessment center, manage it and provide electronic examinations for period of ten years during which some university staff will be trained to take charge of the center after their exit. The eTC centre in FUT Minna consists of two buildings which each house 250 dumb terminals (Wyse V10L Thin Client, value ~USD250 each) and a detached server room and staff offices. The server room houses two separate servers. One of the servers is used to store questions (80GB of hard disk and 4GB of RAM, value ~ USD 1,273) while the second server stores the biometric pictures of the candidates (40GB of hard disk and 2GB of RAM, value ~ USD1,145). The servers and dumb terminals are powered via an uninterrupted power supply. The server room contains a 22002200 American Power ConversionAPC Uninterrupted Uninterruptible Power Supply providing standby electricity in case of occasional power failure, which is still prevalent in the region. Another gadget is 1TB CCTV system that records the activities of candidates during the examination. A 24 port switch is used to connect the systems in the main buildings with the servers.

The software for deploying questions is designed with the interface that authorise the candidate before gaining access to the question using the candidate's identity and registration numbers. The timeframe is attached to the questions and the examination is automatically terminated at the expiration of allocated time. The questions are deployed onto the dumb terminals prior to the commencement of the examination and the candidates can scroll throughout all questions prior to the conclusion of the examination.

Two sample questions are provided below:

1) *Simple Multiple Choice Question*

Who was the vice chancellor of FUT Minna between year 2008 and 2012?
   A. Prof. Tukur Saad
   B. Prof. M. S. Audu
   C. Prof. M. A. Daniyan
   D. Prof. Suleiman Adeyemi

2) *Complex Multiple Choice Question*

   Read the following paragraph and answer the questions below;

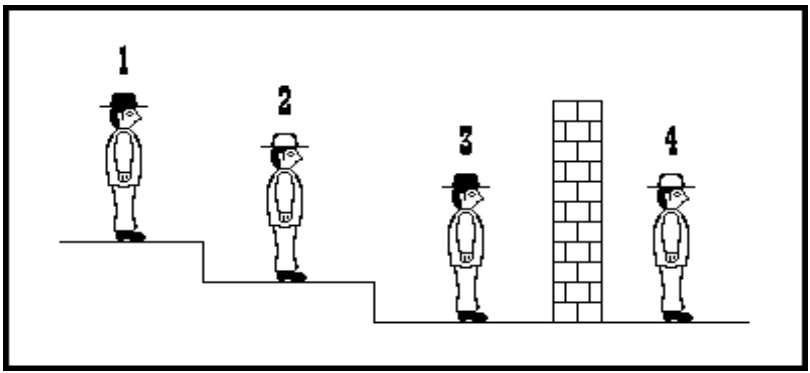

   Figure 1: Source: PHPKB Knowledge Management Software (2012)

Four criminals are caught and are to be punished. The Judge allows them to be freed if they can solve a puzzle. If they do not, they will be hung. They agree. The 4 criminals are lined up on some steps (shown in Figure 1 above). They are all facing in the same direction. A wall separates the fourth man from the other three.

In summary, Man 1 can see Men 2 and 3; Man 2 can see Man 3. Man 3 can see none of the others. Man 4 can see none of the others. The criminals are wearing hats. They are told that there are two white hats and two black hats. The men initially don't know what colour hat they are wearing. They are told to shout out the colour of the hat that they are wearing as soon as they know for certain what colour it is. They are not allowed to turn round or move. They are not allowed to talk to each other. They are not allowed to take their hats off.

Now the aptitude question is "Who is the first person to shout out?

   A. Man 1 will shout first

   B. Man 2 will shout first

   C. Man 3 will shout first

   D. Man 4 will shout first

B.  *eExaminations in Nigeria*

Virtually all examination bodies, academic institutions and professional bodies namely the National Examination Council (NECO), West African Examination Council (WAEC), National Board for Technical Education (NABTEB), Joint Admission and Matriculation Board (JAMB), National Teacher Institute (NTI), Computer Professional Registration Council of Nigeria (CPN) to mention just a few are moving towards engaging eExaminations for the conduct of assessment or registration or both. For example, NECO, WAEC and NABTEB are conducting their registration of candidates online. The year 2013 will see another milestone in the evolution of eExaminations in Nigeria because the JAMB is preparing to start conducting e-testing for its candidates as part of university entry requirements. The computer based test (CBT) is to take place over a period of 17 straight days at different centers across the country. Initially, the eExamination will run in harmony with the old manual test (parallel change over), giving candidates the option to choose either the CBT test or the old manual test; but over several years the old system will be eventually phased-out. A professional body like CPN has been conducting professional examinations for its students and graduate members over time online (www.**cpn**-ncsexam.org).

C.  *University admission in Nigeria*

The university admission in Nigeria is earned based on indices like merit (JAMB score), catchment area, number of applicants' number in afor each course among othersetc.. Due Some poor quality students gain admission due to the high rate of examination malpractices where students they obtain unjustifiable unjustifiably very high scores in JAMB. Such students eventually become a burden to the university. Therefore the universities in Nigeria decided to conduct additional testing of applicants to verify their previous score in the JAMB examination. This is called Post Unified Tertiary Matriculation Examination (PUTME) in various Universities in Nigeria, but due to the high numbers of candidates most institutions prefer electronic examinations and have engaged the services of Electronic Testing Company (eTC), which is a private- public partnership for the exercise.

The eTC center is normally used for PUTME assessments at FUT Minna. The candidates that meet up with the minimum score fixed by JAMB are usually eligible to participate in the assessment. However, each university may also set another standard for their respective institution based on the score index and population of the applicants. Due to the usually large population of these applicants, four or more days are usually earmarked for the exercise, where students are normally grouped based on their intended course of study to ease the conduct of eExaminations and enhance effective management.

The candidates are expected to have previous knowledge of answering online question as no tutorial of any kind is usually given. They are expected to log into the system with their name and identity number. At the end of the examination they are expected to purchase a scratch card (which contains a unique pin code) to check their result just twenty four hours after the examination. The students after twenty hours of the examination can log in to the university portal through their identity and registration numbers and enter their scratch card pin in order to check their results. This very rapid turnaround is a major benefit of the eExamination system provided by eTC.

D. *eExaminations in university courses for FUT Minna*

After admittance to a university course, eExaminations for undergraduate students are mostly compulsory for 100 and 200 level courses (first and second year of study). This is due to the large number of students usually enrolled for some courses (general courses) at those levels, particularly the first year students. However, not all the courses at the 200 level are designated and compulsory for eExamination; the lecturer in charge is given the opportunity to choose an alternative provided the number of students is manageable (not more than 100) and that results will not be delayed unnecessarily. The conduct of these university eExaminations is designed in such a way that all questions are sent to the Deputy Vice Chancellor (academic) for collation. The DVC office then sends the questions to the eTC centre where they are uploaded into the server prior to the examination period. Due to the extensive population of students at these levels, they are usually grouped according to the courses and allocated with different timing in order to allow everybody to write the same examination but at different time. This is because the eTC halls cannot accommodate all the students at the same time. The questions are usually objective and each course is usually allocated a maximum of thirty (30) minutes depending on the nature of the subject. On arrival at the center entrance, the students are screened by security officers to prevent them from entering with unauthorized and/or forbidden materials. They login into the system using their registration number and identity number and must submit their answers before the expiration of their time. The eExamination hall is highly ventilated with about four air conditioning units to provide a cool environment for the dumb terminal equipment (average daytime temperature is 31°C). The examination is also usually supervised by some selected invigilators to monitor and provide normal assistance to the students. The eTC center also has a standby generator and backup power supply for each dumb terminal and server in case of occasional power outages.

E. *Future trajectory in Nigeria*

Apart from FUT Minna's use of eExamination in the conduct of PUTME and undergraduate examinations, [8] attest that at least five other Universities in Nigeria have also adopted the same electronic means of assessing candidates. Presently, most students and lecturers are quite satisfied with the existing system but calls have been made for improvements to cater for the challenges of missing results and security. If the work on biometric security of [8] can be properly implemented, the aforementioned challenges could be drastically reduced or eradicated.

There is no doubt the conduct of eExaminations for PUTME has helped a lot in exposing the atrocity usually committed by candidates. This is not to say that the PUTME is absolutely perfect but it has remained a check on the quality of results being obtained by JAMB candidates. The adoption of eExaminations for some undergraduate courses has also brought about reductions in the challenges being encountered when using manual means of assessment. Due to the large number of students for a particular course, the results might be delayed, missed or even unavailable.

With the acceptance and deployment of eExaminations in Nigerian Universities, there is hope that in the next five to ten years there is going to be a serious improvement in the quality of admitted students and graduates being produced. This might also help to assess the quality of teachings and teaching methodology adopted by the lecturers.

## V. CASE STUDY TWO: UNIVERSITY OF TASMANIA

### A. Background

The eExam System was first used with institutional computers in an information technology course as part of the Bachelor of Education course in 2006. Candidates booted (started up) computers with CD-ROMs containing a live Ubuntu (Linux distribution) in computer laboratories run in shifts to cope with the large number of candidates. Over time, specialized programming efforts modified the live operating system to include anti-collusion features and moved from the CD-ROMs to USB sticks. The strategy behind this innovation was twofold. First, to build a sustainable basis for the necessary equipment (each student provides their own computer for any examination); and secondly to ensure equity of opportunity through providing the identical operating system and application software suite to every candidate. At the time of writing the system has been successfully used by over 1000 candidates in high stakes assessments. The software is open source and can be downloaded from www.eExams.org. The fourth version was in development in late 2012, based upon Ubuntu 12.04 with a design specification which allows both Windows PCs and Macintosh computers to boot from the same USB data stick. The three key security measures implemented provide a fair assessment context by:

- Disabling communications to prevent internet access or collusion
- Interdicting access to local hard drives or other USB data sticks
- Providing a unique visual image (often of a pet animal) on all candidate screens at startup, for non-technical invigilators to check compliance.

Typically a student commencing an eExamination enters the assessment centre with their own computer and finds a USB stick with launching instructions on the desk. The candidate would have downloaded a sample stick at least 3 weeks beforehand, so should be familiar with the process. Having started their computer, the security image is checked before they are permitted to open the question paper and start writing answers. At the end of the exam their answers are left on the USB stick, which is left on the desk for collection, collation and forwarding to the assessors for marking.

### B. University admission sector - Australia

Following presentations of the way the university was moving to local audiences in the school and training sectors, the Tasmanian Qualifications Authority commenced a trial of the eExam System. This authority is responsible for the administration of subject-specific pre-tertiary examinations and the calculation of a national ranking for all students potentially entering university. One subject (information technology and systems) requested to use the eExam System in 2011, and subsequently continued in 2012. A report on the 2011 trial stated

"the e-exam in Information Technology & Systems was done by 93 students at 10 exam centres. Each school had used e-exam for their mid-year examinations and was familiar with the system. There were no issues with major equipment failure" [11]. In 2012 a post-paper question was trialledtrialed by the TQA - see Figure 2.

---

The 'Up and Comers' Tennis Club is a medium-size tennis club with about 150 members. During the consultation process two particular problems were identified:
(i) The tennis club's website is somewhat limited.
(ii) The tennis club's membership records are a mix of:
• some details written on membership forms
• some e-mails received and sent
• some records in a spreadsheet — mainly of those who had paid.
At the time, this was some of the tennis club's website ([click to see it](#)). Specifically referring to this club's website, discuss three aspects of it you would change and why.

*Type your answer below this line:*                                             .

---

Figure 2: Example of a post-paper question from a university admission eExamination, linking to a small web-site provided on the USB stick (no internet connection was allowed or available)

The TQA adoption of the eExam System was done at arm's length, with no assistance provided by the designers after the initial handing over of the code and technical manuals. This is a good demonstration of innovation adoption. Of interest in this context is the implemented definition of equity: all students in the subject using the eExam System are obliged to use a computer. This ensures fairness for all candidates, and eliminates any bias due to differing text entry modalities.

*C.    eExaminations in Tasmanian university courses*

Following the approval for computers to be used in examinations by the academic senate, a range of disciplines have sent faculty staff to training sessions and offered this mode to students as an option or made it a requirement for assessment. Following senate approval the eExam System has been used in History, Law, Education and Medicine, illustrating the diversity of subject applicability. Some of these examinations have been offered to candidates as an option – so they can be described as paper-replacement assessments. Candidates have answered the same questions, but some used the computer keyboard to write their answers, and others have used pens. In other cases, post-paper examinations have been set, so every candidate has required a computer. Examples include Mathematics Education, where a video of classroom practice was provided and candidates asked to criticize the teaching activity portrayed. In 2012 candidates were required to provide their own computer for these assessments, but institutional practice has been to provide a few computers in reserve to cover any shortfall or reliability issues.

See Figure 3 for an example of a paper-replacement question used in the Faculty of Law.

---

In 2011, Alpha Corp was fined $5,000,000 by the Federal Housing Court for collusive pricing under the House Prices Act, 2009 (Cth). It did not appeal the decision to the High Court….
**Provide arguments for the Commonwealth government, including arguments that the High Court should be invited to reconsider the basis of constitutional review.**
Please type your answer below this line

Figure 3: Example of a paper-replacement question from the Bachelor of Laws eExamination.

Notice that in this context the principle of equity is associated with the opportunity to choose a candidate's preferred writing implement (pen or keyboard). Also that post-paper examinations including digital materials such as video clips and instructional software are already being set.

*D.     Future trajectory in Australia*

Over the next 5-10 years the eExam System is expected to be developed so that less technical expertise is required by candidates. Some of the challenges are the move from the BIOS system to EFI as Windows 8 is deployed. Greater numbers of ARM processors replacing the Intel classes may need to be accommodated, and the popularity of tablets is under consideration. Typically these latter do not provide ways of booting from an alternative operating system source, and are arguably poor devices upon which to input text at high speed (typical of examination conditions).

Duplication of USB sticks is a labour-intensive activity which may be replaced by a network reticulation system. Hitherto the use of communications has been avoided as part of a design strategy to eliminate single-point failures; the hardware failure of a single wireless access point could disrupt the examination of several hundred candidates. Also, operational communications links could conceivably be subverted by candidates breaching security protocols.

## VI.    RESULTS

A comparative analysis of these two institutions shows that both systems are becoming accepted as part of assessment in their individual countries. In Nigeria, there is evidence of widespread systemic adoption for university entrance merit selection, whilst in Australia this has been confined to one subject in one state. Within undergraduate courses, the Nigerian experience appears to be quite extensive in one university (all first year courses), while in Australia this adoption has been slower but has penetrated a wide variety of disciplines.

Another comparison looks at the nature of assessment in the two institutions. The Nigerian case study shows candidates are limited to selecting from a list of prepared answers to each question. This allows automatic marking and provides an extremely fast turnaround time for assessment results. The Australian technology supports essays (marked conventionally by a human assessor) and sophisticated software use within the identical operating system environment for all candidates.

The result of the analysis of the existing eExamination system in Nigeria and Australia is to have a form of USB stick-based eExamination that is deployable on the internet with encryption and decryption features using public key cryptography, which allows lecturers to encrypt their questions and post them online while the questions is decrypted by the administrator or students given the key. The system shall also be essay and objective question oriented.

Table 1: A Comparative Table

| eExamination Features | FUT Minna, Nigeria | University of Tasmania, Australia |
|---|---|---|
| Type of questions | Objective questions only | Both essay and objective questions, and full use of the Ubuntu operating system within security constraints. |
| Population size at a time | A maximum of about 500 students can sit for the exams at a time. | The largest cohort assessed has been 120 candidates in a single hall, 150 candidates spread across multiple testing sites, including overseas locations (simultaneously). |
| Technologies used and proposed | <ul><li>Administered within a LAN or intranet environment</li><li>A server and a student information database are connected to the LAN.</li><li>A question bank in the database was also connected to the intranet.</li></ul> | <ul><li>Candidates' individual systems boot from USB sticks.</li><li>Ubuntu 12.04 with a design specification which allows both Windows PCs and Macintosh computers to boot from the same USB data stick.</li><li>Questions come from individual lecturers, and can either be paper-replacement or post-paper in design.</li></ul> |
| Security Features | <ul><li>Username and passwords for both students and administrative login.</li><li>Cryptographic technique have been proposed to secure the database</li><li>Biometric authentication to curb impersonation.</li></ul> | <ul><li>Disabling communications to prevent internet access or collusion</li><li>Interdicting access to local hard drives or other USB data sources</li><li>Providing a unique visual image (often of a pet animal) on all candidate screens at startup, for non-technical invigilators to check compliance.</li></ul> |

Table 1 above shows a summary of the comparison of the features of the eExamination systems we considered in the two case studies in this research work.

VII. CONCLUSION

In our opinion the prospects for more extensive use of eExaminations is very good for a wide range of disciplines. This is not to say they will be used in all disciplines, and some subjects may be more suited to the technique than others.

The contrast between the two universities suggests there may be a fruitful convergence of the approaches. Dedicated buildings and equipment put considerable strain on resources, which may not be sustainable without public-private partnerships. However, the ease of deployment using a web-centred strategy is one which overcomes the difficulties associated with preparing and distributing USB sticks.

Therefore a future design may use a web-based strategy to prepare a virtual machine which is launched into candidate computers at the start of each examination. This would retain the benefits of having an entire operating system for the assessment period, and ensure every candidate has precisely the same software tools. However, it would also move the examination beyond knowledge-recall to assess higher level professional skills.

At the same time, security needs will be paramount, and will possibly incorporate biometric and personal supervision to ensure fairness. The days when a candidate can be tested in their own home without a trusted person to assure their identity and wholly personal responsibility for all the answers has not yet come!

The comparative analysis of the eExamination systems in both countries has therefore helped the authors devise a new eExamination system that contains the best features of systems in both countries. The new system is a form of USB stick-based eExamination (Australia) that is deployable on the internet with encryption and decryption features using public key cryptography (Nigeria proposed system). This will allow lecturers to encrypt their questions and post them online while the questions are decrypted by the administrator or students given the key. The system shall also be essay and objective question oriented.

**Authors**

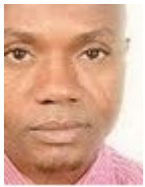

Olawale Surajudeen Adebayo (MCPN, MNCS) is a Lecturer in the Department of Cyber Security Science, Federal University of Technology Minna, Niger State, Nigeria. He bagged B.Tech. in Mathematics and Computer science from Federal University of Technology, Minna in 2004 and the MSc. in Computer science from University of Ilorin, Kwara State, Nigeria in 2009. He is presently a PhD student in the Department of Cyber Security Science, Federal University of Technology, Minna. His current research interests include: Information Security, Cryptology, and Data Mining Security. He has published some academic papers in the above-mentioned research areas within and outside Nigeria. He is a member of Computer Professional Registration Council of Nigeria (CPN), Nigeria Computer Society (NCS), Global Development Network, International Association of Engineers (IAENG) and many others. He is a reviewer to more than five local and international journals.

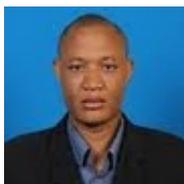

**Shafi'i Muhammad ABDULHAMID** holds M.Sc. degree in Computer Science from Bayero University Kano, Nigeria (2011) and B.Tech. degree in Mathematics/Computer Science from the Federal University of Technology Minna, Nigeria (2004). His current research interests are in Software Engineering, Cyber Security, Computational Intelligence and Operating Systems. He has published many academic papers in reputable journals within Nigeria and Internationally. He is a member of International Association of Computer Science and Information Technology (IACSIT), International Association of Engineers (IAENG), The Internet

Society (ISOC) and a member of Nigerian Computer Society (NCS). Presently he is a lecturer at the Department of Cyber Security Science, Federal University of Technology Minna, Nigeria.

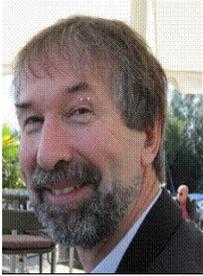

Dr. Fluck has taught high school mathematics, science and information technology in Nigeria, England and Australia. He has been a teacher educator at the University of Tasmania since 1995 where his focus has been on the transformational potential of computers in education. He has provided advice on the emerging national curriculum in Australia; developed an eExam system for students to use their own computers in high stakes assessment; and serves on the executive of Working Group 3.3 (research into educational applications of information technologies) for IFIP/UNESCO. He has published widely on information technology in education, designed a chair and is a member of the Paringa Archers club in Launceston. Andrew is partnered with three excellent children. See more at http://www.educ.utas.edu.au/users/afluck/.

Andrew is now working as a Senior Lecturer of Information Technology in the Faculty of Education at the University of Tasmania. We have a large server which provides staff and accounts, web-services, and numerous other services. Andrew first visited Tassie on a lecture tour in 1998, and moved there the following year. His career has included spells in the bush of Biafra, the social forefront of Milton Keynes and the cutting edge of special education in southern England. With his partner Filomena, and three kids, he knows the value of peace & tranquility.